\documentclass[ACS,STIX2COL]{WileyNJD-v2}
\articletype{ORIGINAL ARTICLE}

\usepackage{mathtools}
\usepackage{graphicx}
\usepackage{dcolumn}
\usepackage{array}
\usepackage{lipsum}
\usepackage{bm}
\usepackage{subfigure}
\usepackage{multirow}
\usepackage{tabularx}
\usepackage{amsmath}
\usepackage{braket}

\newcommand{\A}{\textup{\AA}}

\received{xxx}
\revised{xxx}
\accepted{xxx}

\usepackage{bm}

\begin{document}

\title{Orientational Effects in the Low Pair Continuum of Aluminium
}

\author[1,2]{Thomas Gawne*}
\author[1,2]{Zhandos A. Moldabekov$^\dagger$}
\author[4]{Oliver S. Humphries}
\author[4]{Motoaki Nakatsutsumi}
\author[1,2]{Sebastian Schwalbe}
\author[2]{Jan Vorberger}
\author[4]{Ulf Zastrau}
\author[1,2]{Tobias Dornheim}
\author[4]{Thomas R. Preston$^\S$}

\authormark{Thomas Gawne \textsc{et al}}

\address[1]{\orgdiv{}\orgname{Center for Advanced Systems Understanding (CASUS)}, \orgaddress{D-02826 \state{G\"orlitz}, \country{Germany}}}

\address[2]{\orgdiv{} \orgname{Helmholtz-Zentrum Dresden-Rossendorf}, \orgaddress{ D-01328\state{ Dresden}, \country{Germany}}}

\address[4]{\orgdiv{}\orgname{European XFEL, Germany},  \orgaddress{D-22869  \state{Schenefeld}, \country{Germany}}}

\corres{*\email{t.gawne@hzdr.de}\\
$^\dagger$\email{z.moldabekov@hzdr.de}\\
$^\S$\email{thomas.preston@xfel.eu}}

\authormark{Thomas Gawne \textsc{et al}}

\abstract{
We compare the predictions of the dynamic structure factor (DSF) of ambient polycrystalline aluminium from time-dependent density functional theory (TDDFT) in the pair continuum regime to recent ultrahigh resolution x-ray Thomson scattering measurements, collected at the European XFEL. TDDFT predicts strong anisotropy in the DSF at the wavenumber examined here, even with $q$-blurring accounted for. The experimental spectrum has more than sufficient resolution and signal-to-noise levels to resolve these orientation dependencies, and therefore the orientational averaging of the polycrystalline sample is observed rigorously. Once the orientation averaging is accounted for, TDDFT is able to reproduce the experimental spectrum adequately. Finally, comparisons of predicted DSFs from jellium to experiment demonstrates the importance of accounting for lattice effects in modelling the spectrum from a polycrystal.
}


\maketitle

\section{Introduction}\label{sec:intro}

X-ray Thomson scattering (XRTS)~\cite{sheffield2010plasma, siegfried_review} is a powerful and commonplace technique to study matter in a wide range of conditions. By scattering x-rays off the electrons in a system, the energy loss (and gain) spectrum of the photons reveals the electronic dynamic structure factor (DSF) of the system~\cite{Dornheim_review,Gregori_PRE_2003}. This spectrum can then be used to infer system conditions (e.g. temperature~\cite{DOPPNER2009182,Dornheim_T_2022,Dornheim_T2_2022,Mo_PRL_2018} and density~\cite{Tilo_Nature_2023,dornheim2024unraveling,dornheim2024modelfreerayleighweightxray,dharmawardana2025xraythomsonscatteringstudies, moldabekov2024ultrafast}), as well as electronic structures and correlations within the system~\cite{kraus_xrts,Witte_POP_2018,Dornheim_SciRep_2024, Moldabekov_2024_Excitation}.
Furthermore, the scattering vector $\bm{q}$ is inversely proportional to the probing length scale, meaning that by simply changing the position of a detector, it is possible to probe different properties in the system; i.e. low scattering angles probe collective properties, and high scattering angles probe individual particles~\cite{bellenbaum2025estimatingionizationstatescontinuum}.

To extract information from XRTS measurements, the standard approach is to fit a DSF model to the spectrum (while accounting for broadening by the source-and-instrument function (SIF)~\cite{Gawne_2024_SIF}), but this means the inferred conditions are then sensitive to the underlying assumptions made in the DSF model~\cite{boehme2023evidence,Gregori_PRE_2003,Sperling_PRL_2015,Ramakrishna_PRB_2021,Mo_PRL_2018}. There is therefore great interest in understanding the accuracy of theoretical approaches for calculating the DSF in different scenarios~\cite{dynamic2,Takada_PRB_2016,Panholzer_PRL_2018,Hentschel_pop_2023,Schoerner_PRE_2023,Bonitz_pop_2024,moldabekov_MRE_2025,White_2025,chuna2025dualformulationmaximumentropy}.

Popular DSF codes that are used to extract conditions from XRTS measurements are based on the Chihara decomposition~\cite{Chihara_JoP_1987,Chihara_JoP_2000,Gregori_PRE_2003}, which categorises the electrons as either bound or free. The free electrons are then typically treated as a uniform electron gas~\cite{review} (jellium), either at the level of the random phase approximation (RPA) or with the local field correction (LFC)~\cite{quantum_theory,dornheim_dynamic,Dornheim_PRL_2020_ESA,Fortmann_PRE_2010}. However the agreement of these models with experimental data can be poor, particularly when there is a strong influence from the ions on the DSF. 
A more detailed approach to calculate the full DSF uses linear response time-dependent density functional theory (LR-TDDFT)~\cite{ullrich2012time,marques2012fundamentals,Moldabekov_prr_2023, moldabekov_MRE_2025, Moldabekov_jcp_2023}, which self-consistently calculates the electronic response to a perturbation in the full electronic and ionic environment.

Recently, ultrahigh resolution XRTS measurements have been used to benchmark the predictive capability of TDDFT in polycrystalline aluminium (Al)~\cite{Gawne_2024_Ultrahigh} and single crystal silicon (Si)~\cite{Gawne_2025_Si}, both in ambient conditions. In both cases, TDDFT was able to reproduce the observed XRTS signal once the finite size of the spectrometer collecting optic was accounted for. In the case of the Al data, the comparison was limited to the spectra measured in the collective (plasmon) regime.
For these spectra, the inelastic signal was found to be largely isotropic with respect to the orientation of the scattering vector through the lattice.
However, an additional spectrum was also measured at a scattering vector of $q=1.73~\A^{-1}$, which is well above the critical wavevector ($q_c\sim1.346~\A^{-1}$) of Al. This spectrum therefore lies in the electron-hole pair continuum~\cite{quantum_theory}, where the plasmon decays into a multitude of excitations due to Landau damping.
For this scattering vector, a single orientation was unable to model the measured spectrum.
However, earlier work in the non-collective regime (at $q=2.8~\A^{-1}$) has suggested orientational effects in the inelastic spectrum become important and need to be accounted for~\cite{Larson_2000_Correlation}.

Here, we examine the contribution of the different orientations of the crystallite domains in the polycrystalline Al to its scattering spectrum at a scattering vector between those of recent measurements in Al~\cite{Gawne_2024_Ultrahigh} and earlier work at synchrotrons~\cite{Sternemann_2000_Effect, Larson_2000_Correlation}. 
Whilst we find that $q$-blurring has a noticeable effect on the shape of the spectra along a given direction like previous works~\cite{Gawne_2024_Ultrahigh, Gawne_2025_Si}, it is not the only effect, and some excitations are only weakly affected by the blurring.
We also find from TDDFT that the inelastic scattering spectrum is strongly orientation-dependent at this wavevector, and that orientation averaging needs to be accounted for in the comparison of theory to experiment.
Only once $q$-blurring and orientation averaging are accounted for, do we find that TDDFT reproduces the shape of the spectrum quite well. Lastly, we compare our results to predictions from the RPA and LFC models of jellium. Neither are capable of reproducing the experimental data, demonstrating the importance of accounting for ionic and lattice effects when modelling (ambient) crystalline systems, including those which are polycrystalline.

\section{Methods}\label{sec:methods}

The scattering spectrum shown here was collected at the European XFEL in Germany~\cite{Zastrau2021} as part of the experiment presented in Ref.~\cite{Gawne_2024_Ultrahigh}; a detailed description of the experimental setup is provided there.
The authors observed that by probing the DSF using a monochromated XFEL beam, and collecting the data on a high resolution spherically-bent Si (533) diced crystal analyser (DCA)~\cite{Descamps_SR_2020,Wollenweber_RSI_2021}, an experimental resolution of $\Delta E \sim 0.1$~eV was achieved, and the broadening by the source-and-instrument function (SIF) of the experiment does not substantially affect the shape of the measured spectrum.
The greatest source of experimental broadening comes from the finite size of the DCA, meaning it covers a range of scattering angles. As a result, the measured spectrum is an average over the range of scattering vectors covered by the spectrometer, an effect known as $q$-blurring or broadening. To limit this effect, a horizontal slit mask was placed in front of the DCA to limit the angular coverage to $\pm1.4^\circ$. Nevertheless, analysis of the data collected using this setup has shown that it is still important to account for this effect in theoretical calculations~\cite{Gawne_2024_Ultrahigh,Gawne_2025_Si}. Specifically, a meaningful comparison of theory to experiment was achieved by averaging a number of theoretical DSFs with scattering vectors in the coverage of the detector for all scattering wavevectors below the critical wavevector.
The spectrum analysed here was collected from a 50~$\mu$m thick free standing polycrystalline Al foil at the largest scattering angle of $25.6\pm1.4^\circ$, corresponding to a scattering vector range of $q=1.730\pm0.095~\A^{-1}$. Due to the low beam energy, heating by the XFEL is negligible, so the sample was probed at ambient conditions.

The Liouville-Lanczos approach to linear response TDDFT, implemented in Quantum ESPRESSO \cite{Giannozzi_2009, Giannozzi_2017, Giannozzi_jcp_2020, Carnimeo_JCTC_2023, TIMROV2015460}, was used for the calculations of the DSF for Al with a face-centered cubic (fcc) structure. We used a $20\times20\times20$ $k$-point grid and an energy cutoff of $75~{\rm Ry}$. A cubic simulation cell was used with a lattice parameter of $a=4.05~\A$~\cite{wyckoff1948crystal}. The results were computed using the Lorentzian smearing parameter  $\eta=0.5~{\rm eV}$. The number of Lanczos iterations was set to $N_{\rm iter}=10^4$. After $N_{\rm iter}$ iterations, we used a bi-constant extrapolation scheme to compute $10^5$ Lanczos coefficients. The used pseudopotential {Al.pbe-rrkj.UPF} is from the Quantum ESPRESSO pseudopotential database~\cite{PP_ESPRESSO_2022}. We used the PBE exchange-correlation functional \cite{PBE_1996}.
Calculations were performed with the scattering vector aligned along the X [100], K [110], L [111], W [120], and U [141] directions; i.e. the five directions connecting the $\Gamma$ point to the high symmetry points of the Brillouin zone (BZ). To account for $q$-blurring, for each of these orientations five LR-TDDFT calculations were performed at wavenumbers 1.635, 1.6825, 1.73, 1.7775, 1.825~$\A^{-1}$ within the angular coverage of the DCA. 
The atomic structure and the components of the scattering wavevectors relative to the simulation cell were defined using the Atomic Simulation Environment \cite{Hjorth_Larsen_2017}.

For the orientation averaging, the DSFs along each direction are weighted by the area in reciprocal space nearest to each central scattering vector along the surface of the BZ. These weights (normalised by $\pi/\sqrt{2}a$) for each direction are: X: 1/6, K: $\sqrt{3}$/6, L: $\sqrt{3}$/3, W: 1/6 + $\sqrt{3}$/3, and U: 1/6 + $\sqrt{3}$/6. The total area is $1/2 + \sqrt{3}$.

\section{Results}\label{sec:results}
\begin{figure*}
    \centering
    \begin{subfigure}
        \centering
        \includegraphics[width=0.45\linewidth,keepaspectratio]{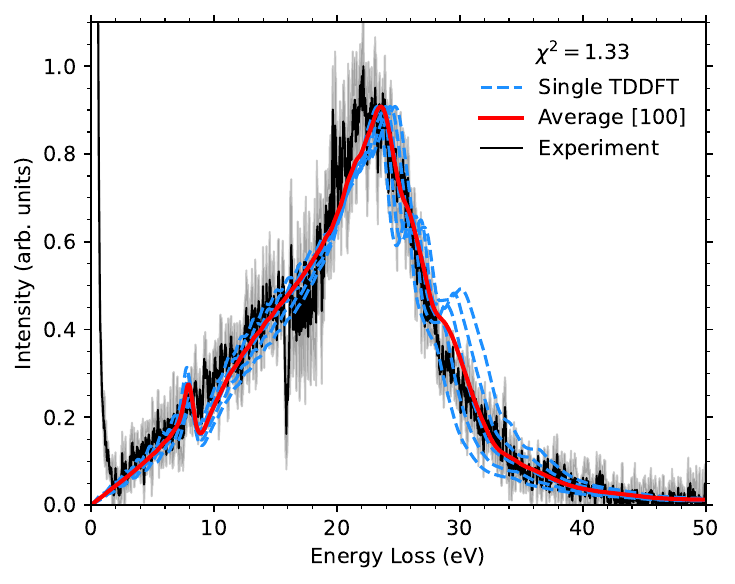}
    \end{subfigure}
    ~
    \begin{subfigure}
        \centering
        \includegraphics[width=0.45\linewidth,keepaspectratio]{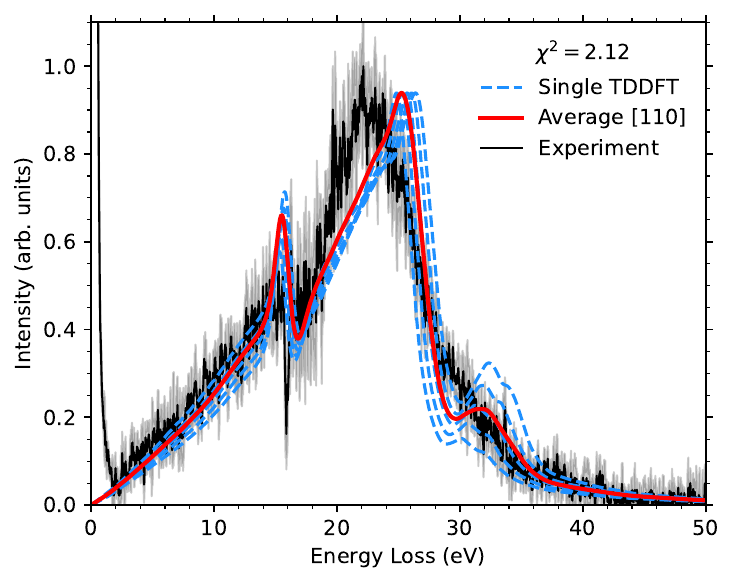}
    \end{subfigure}
    \vspace{0.1mm}
    \begin{subfigure}
        \centering
        \includegraphics[width=0.45\linewidth,keepaspectratio]{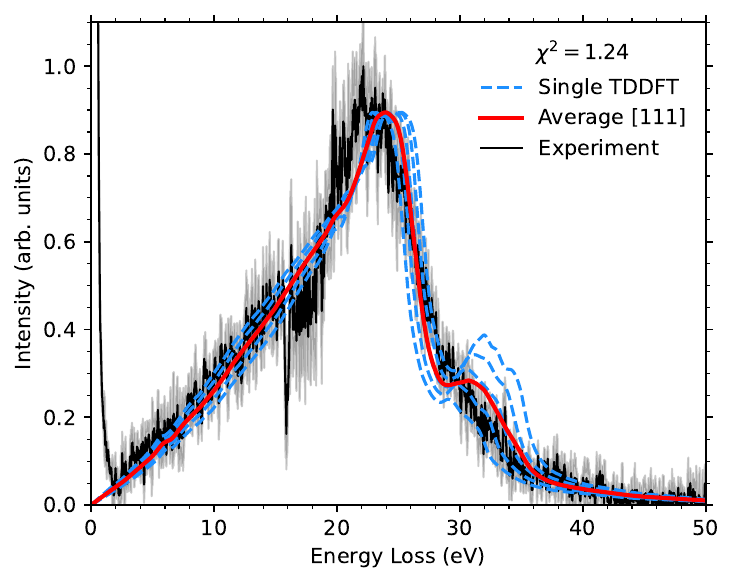}
    \end{subfigure}
    ~
    \begin{subfigure}
        \centering
        \includegraphics[width=0.45\linewidth,keepaspectratio]{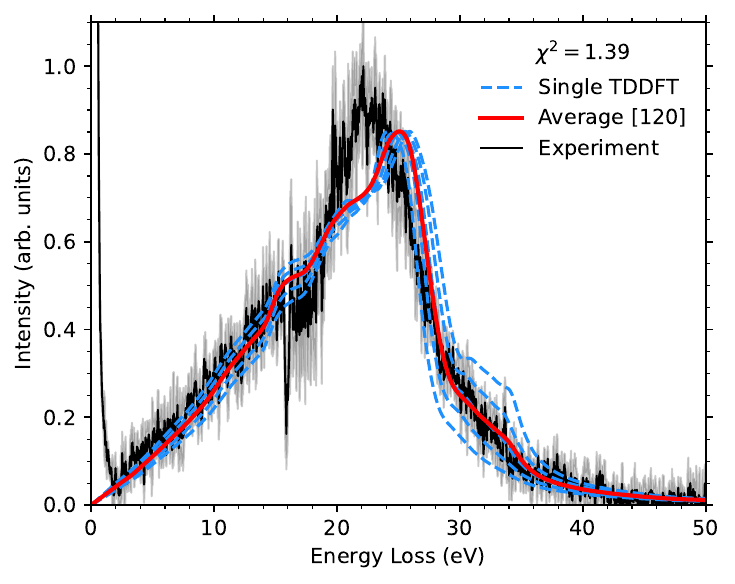}
    \end{subfigure}
    \vspace{0.1mm}
    \begin{subfigure}
        \centering
        \includegraphics[width=0.45\linewidth,keepaspectratio]{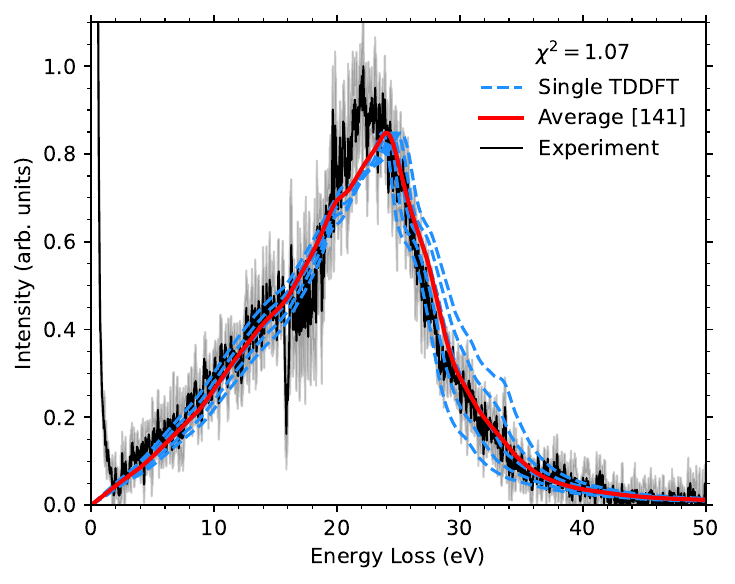}
    \end{subfigure}
    ~
    \begin{subfigure}
        \centering
        \includegraphics[width=0.45\linewidth,keepaspectratio]{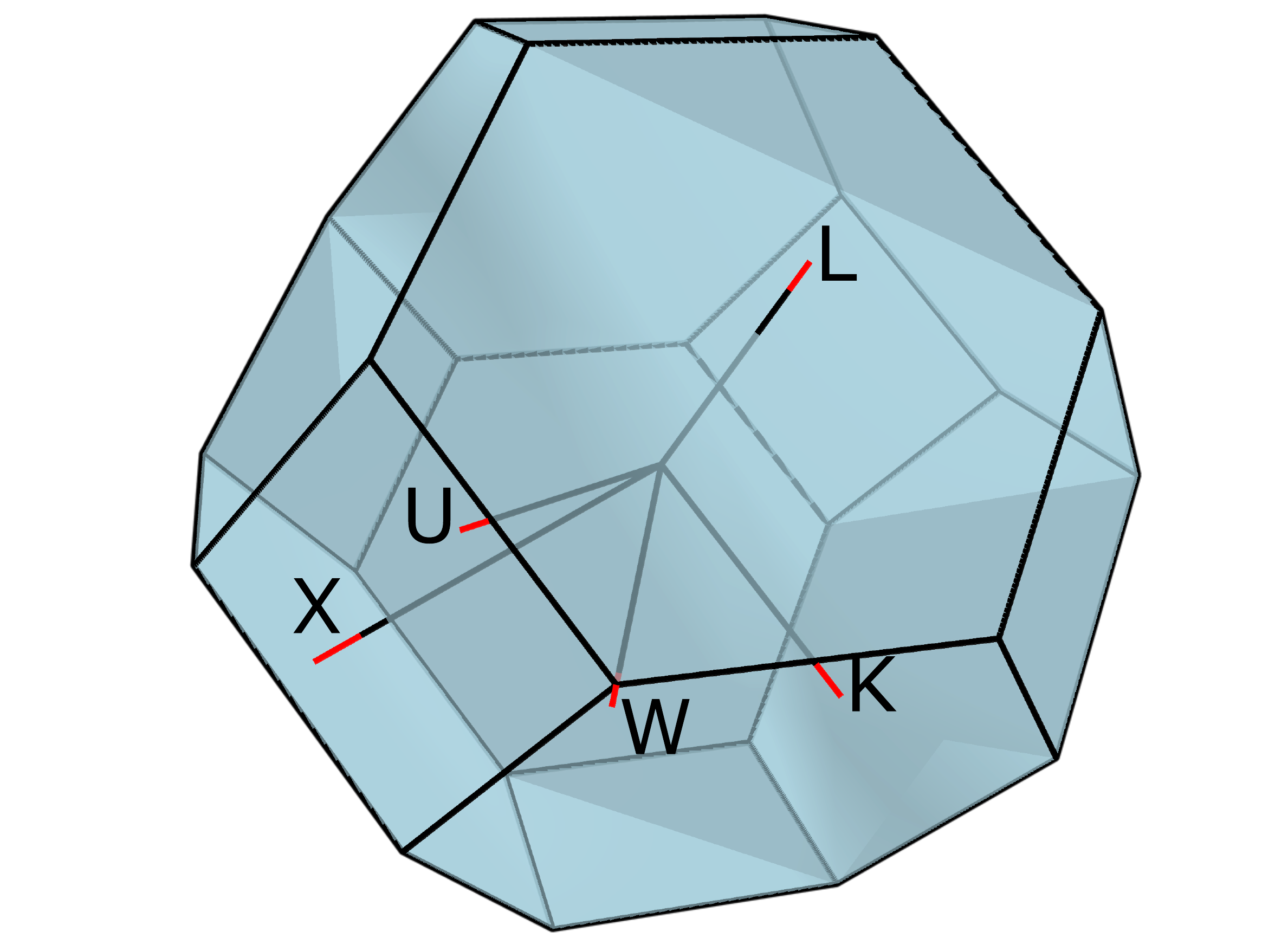}
    \end{subfigure}
    \caption{Comparison of TDDFT calculations of the DSF along the five principle directions (X [100], K [110], L [111], W [120], and U [141]) to the experimental spectrum collected at scattering vectors in the range of $q=1.730\pm0.095~\A^{-1}$, shown by the black solid line. Single TDDFT calculations are shown by the blue dashed lines, which are uniformly distributed in the $q$-range (with increasing $q$ from left to right). The average of these curves is shown in the red solid line. Each of the TDDFT DSFs are normalised to the same maximum as the average DSF, however for the averaging they were not normalized beforehand. The reduced-$\chi^2$ values in the plot are for the averaged TDDFT curve to the experimental data. The bottom right plot shows the BZ of Al and the principle directions. The red portion of the lines indicates the $q$-vector range covered by the detector along each direction.
    }
    \label{fig:Principal}
\end{figure*}

To understand the importance of orientational averaging, we investigate the degree of anisotropy in the DSF predicted by TDDFT, and to what degree it is affected by $q$-blurring. In Fig.~\ref{fig:Principal} we show the DSF along the five principle directions directions for wavenumbers within the detector coverage, along with the uniform average of these DSFs which models the $q$-blurring effect.
Evidently, the DSF predicted by TDDFT is very anisotropic at this wavenumber, even with $q$-blurring accounted for. While some prominent features in the spectra are dampened by the $q$-blurring -- for example, the narrow excitations seen around 30~eV in all orientations -- other features, such as the spike at $\sim15$~eV in the [110] DSF and $\sim8$~eV spike for [100], are not heavily suppressed.
This 8~eV spike has been previously observed in an inelastic x-ray scattering experiment in single crystal Al at the Fermi wavenumber along the [001] direction~\cite{Tischler_2003_Interplay}, which is orientationally equivalent to scattering along the [100] direction. Likewise, this feature has been predicted by ALDA TDDFT along the [100] direction in the non-collective regime ($q = 1.55$--1.94~$\A^{-1}$)~\cite{Tischler_2003_Interplay,Cazzaniga_PRB_2011}. Taken together, we therefore have confidence in the orientational predictions of TDDFT calculations presented here.
That these low energy wing spikes in the [100] and [110] DSFs are clearly not observed in the experimental data here indicates that the suppression of these features will be predominantly due to orientational averaging.

For all five orientations, we see, to varying degrees, minor discrepancies between the experimental spectrum on the high energy wing due to the shape of the additional excitation in the theoretical DSFs.
Since the resolution afforded by the experimental setup is very high, and the signal-to-noise ratio of the measurement is good, if the predictions of TDDFT are accurate then the signatures of any dominant orientation would be observable, which was the case for single crystal Si measurements using the same setup~\cite{Gawne_2025_Si}.

\begin{figure}
    \centering
    \includegraphics[width=\columnwidth,keepaspectratio]{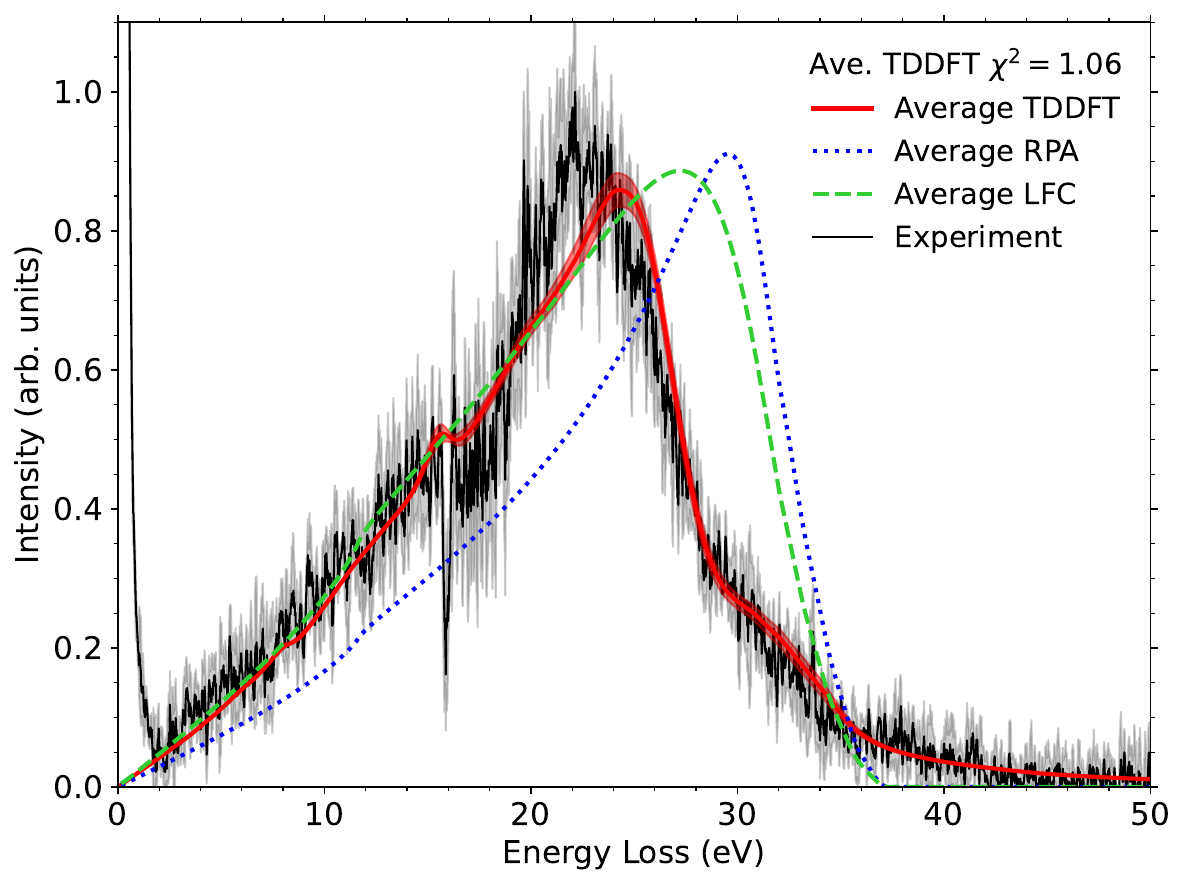}
    \caption{Comparison of different theoretical DSFs to experiment (black): average of the $q$-blurred DSFs from TDDFT shown in Fig.~\ref{fig:Principal} (red), and the $q$-blurred DSF for the uniform electron gas using the random phase approximation (blue dotted) and local field correction (green dashed). The red area indicates the standard error of the averaged TDDFT calculations over all orientations and q-vectors. The dip intensity in the experimental spectrum at 15.8~eV corresponds to a detector gap, and is otherwise unphysical. The reduced-$\chi^2$ value between the average TDDFT curve and the experimental spectrum is indicated in the figure.
    }
    \label{fig:Average}
\end{figure}

Since the target is polycrystalline Al, the measured spectrum is in reality an average of the scattering spectra from different grain orientations in the sample. In Fig.~\ref{fig:Average}, we average the five $q$-blurred TDDFT DSFs to produce an orientation-averaged spectrum. To account for the fact that the BZ of Al is a truncated octahedron, and so a $\bm{q}$ passing through a randomly oriented grain is more likely to pass near some high symmetry directions than others, the DSFs are weighted by the area in reciprocal space nearest to their respective directions.
As should be expected, the prominent orientation-specific features on the low energy wing are greatly suppressed in the orientation-averaged DSF.
Due to the suppression of the orientation-specific low energy wing excitations by the contributions from the other two orientations, the agreement of this averaged spectrum to the experimental data is substantially improved compared to most of the directions, indicated by the reduction in the reduced-$\chi^2$ parameter, and a minor improvement by this measure over the [141] direction. 
Furthermore, the orientation averaging of the excitation around 30~eV results in better agreement of the high energy wing from TDDFT with experiment than the individual spectra as well.
There are however still some differences between the TDDFT and the experimental spectra. For example, TDDFT under-predicts the intensity between 21.2--22.8~eV, leading to a slightly higher ($\sim1.45$~eV) peak position than inferred from the experiment~\cite{Gawne_2024_Ultrahigh}. Additionally, there is still a small spike at 15~eV from the [110] DSFs, though it is now substantially suppressed by the orientation averaging.

Some of the differences may be ascribed to the chosen pseudopotential, and a comparison for the DSF along the [120] direction is shown in the Appendix in Fig.~\ref{fig:Psuedo}. While there are some minor differences in the shapes of the wings, the peak position is consistent between pseudopotentials. It is worth noting that the pseudopotential is a chosen input into DFT that does affect predicted properties. Yet, pseudopotentials are generally not optimised to give the best DSF, but rather other properties such as the atomisation energy~\cite{Lejaeghere_2014_Error}, so it is not surprising that the choice of pseudopotential has some effect on the shape of the DSF (see e.g. Ref.~\cite{moldabekov_MRE_2025}). The theoretical spectrum could therefore be in principle ``optimised'' by trialling different pseudopotentials, however given the scale of the differences observed in Fig.~\ref{fig:Psuedo}, this is unlikely to substantially change the discrepancies seen in Fig.~\ref{fig:Average}.
Nevertheless, we can conclude that orientational averaging can be an additional important consideration in modelling the DSF of crystalline systems, especially those in which the DSF depends on the specific orientation through the lattice.

The results presented here contrast with the measurements of Al in the collective regime~\cite{Gawne_2024_Ultrahigh}. There, TDDFT predicted that the DSF was relatively isotropic along the principal reciprocal lattice vectors, so the experimental spectra could be reproduced simply by the DSF along the [100] direction, once $q$-blurring was accounted for.
In another earlier study~\cite{Larson_2000_Correlation}, orientational effects in polycrystalline Al were examined at a higher wavenumber of $q = 2.8~\A^{-1}$ using TDDFT. There, the DSF was calculated for ten directions inside the triangle between the [100], [110], and [111] directions, and then averaged. While some differences can be observed between the different orientations, they are not nearly as stark as observed here in Fig.~\ref{fig:Principal}, which may be attributed to the larger wave vector; it is also noted there that the measurements~\cite{Sternemann_2000_Effect} of the inelastic spectrum in polycrystalline Al at $q = 2.8~\A^{-1}$ are extremely similar to the spectrum along the [013] direction in single crystal Al, which suggests lattice effects are overall less important due to the smaller probing length scales at high scattering vectors. Nevertheless, they observe a smoothing effect from orientation averaging even at the higher wavevector.

To emphasise this point further, the BZ of Al lies between $0.67 - 1.736~\A^{-1}$ (the distances $\Gamma L - \Gamma W$ in reciprocal space) meaning that anisotropy in the Fermi surface of Al can be uniquely probed by x-rays with wavevector $q=1.73~\A^{-1}$. The electrons are sited around the edges of square face of the BZ~\cite{Harrison1959, Ashcroft1963} between the $U$ and $W$ points which are found between $1.643 - 1.736~\A^{-1}$ ($\Gamma U - \Gamma W$), and similarly, towards the K point which has the same distance $\Gamma U$. The free electrons form a connected surface in these directions in reciprocal space~\cite{Harrison1959}.

Finally, we compare our results to predictions of the DSF from jellium using both the random phase approximation (RPA) and jellium model with the local field correction (LFC) from quantum Monte Carlo simulations \cite{PhysRevB.103.165102, Dornheim_PRL_2020_ESA,dornheim_ML}. Since Al is a prototypical free-electron-gas metal, jellium models are often considered a good candidate for modelling its dynamic response. However, at very low wavenumbers, the jellium prediction of the plasmon shift in Al is too large since it neglects the localization of the free electrons due to the presence of the ions. At high wavenumbers, the predictions are more accurate since electrons are probed on smaller length scales (eventually at the level of single electrons), such that the effect of the ions on the free electrons is no longer measurable.
Previous inelastic x-ray scattering measurements in polycrystalline Al at $q = 2.8~\A^{-1}$ were compared to jellium models~\cite{Sternemann_2000_Effect,Larson_2000_Correlation}. There, reasonable agreement can be seen between the theoretical and experimental spectra when LFC is included, but there were still some notable discrepancies. No agreement was found at the level of RPA.
The scattering vector in this work lies in the pair continuum, but also in between the collective regime and the earlier higher scattering vector measurements in Ref.~\cite{Sternemann_2000_Effect}.
We therefore evaluate the predictive capability of RPA and LFC at this intermediate wavevector, which is also shown in Fig.~\ref{fig:Average}.
Both the RPA and LFC DSFs here have $q$-blurring included, which results in broadening of the DSFs. As expected, we find RPA is unable to reproduce the shape of the scattering spectrum, nor does it produce an accurate estimate of the position of the maximum of the experimental data.
The spectrum calculated with LFC is also poor: while it produces a somewhat better estimate for the lower energy wing below 21~eV, the position of the maximum remains substantially higher than experiment.
For both RPA and LFC, the high energy wing decays much more sharply than observed in experiment, and are positioned far from the experimental measurement.

\section{Discussion and conclusions}\label{sec:conclusion}

We have presented a first-principles analysis of ultrahigh resolution XRTS data from polycrystalline Al collected in the electron-hole pair continuum region. In contrast to previous analysis of XRTS measurements in the collective regime~\cite{Gawne_2024_Ultrahigh}, here we find there is a strong dependence of the DSF on the orientation of the scattering vector through the crystal, even with $q$-blurring accounted for. In other words, the scattering vector extends well beyond the first BZ, and is therefore sensitive to the crystal orientation. The anisotropy of the DSF is also much larger compared to previous TDDFT calculations at a large scattering vector~\cite{Sternemann_2000_Effect}, which we attribute to the probing length scale matching the reciprocal lattice vectors.

Due to the polycrystalline nature of the sample, and the strong anisotropy of the DSF, we find it is necessary to account for orientation when comparing TDDFT results to the experiment.
While $q$-blurring~\cite{Gawne_2024_Ultrahigh,Gawne_2025_Si} does flatten features on the high energy wing of the DSFs, orientation-specific low energy wing excitations are are largely unaffected by $q$-blurring.
Once the orientational averaging is accounted for, we find TDDFT is able to make reasonably good predictions of the scattering spectrum in polycrystalline Al in the non-collective regime, although there are some discrepancies such as the shape of the DSF around the maximum. In contrast, both RPA and LFC are unable to make good predictions of the shape of the DSF, nor its peak position, even with $q$-blurring accounted for.

We therefore concluded that a full treatment of the linear response of the electronic structure in the presence of ions is necessary to make accurate predictions of the shape of the DSF at low scattering angles in polycrystalline Al.
This has potential significance for XRTS codes that utilise RPA and LFC models to estimate the free-free response of a system, particularly in cases where the codes are used to infer system conditions from XRTS spectra, including in cases where the system contains a crystalline ionic structure.
While the sample here is at ambient conditions, we expect that the strong effect of the lattice on the shape of DSF will continue into isochorically heated systems as the electronic structure is still influenced by the presence and positions of the ions.
However, we do not draw further conclusions on the applicability of RPA or LFC to disordered systems: while the effects of the crystal structure are smoothed due to the polycrystalline nature of the sample, it is evident that the crystalline structure of the domains is still prominent. Therefore, lattice and localization effects are still present and have a substantial effect in the present data. Further investigation is therefore required in order to comment on the applicability of RPA and LFC to disordered systems.

\section*{Data Availability}

The experimental data supporting these results have been reused with permission, and can be found here and are available upon reasonable request: doi:10.22003/XFEL.EU-DATA-003777-00. Simulation data supporting these results is available here: doi:10.14278/rodare.3907.

\section*{Acknowledgements}

This work was partially supported by the Center for Advanced Systems Understanding (CASUS), financed by Germany’s Federal Ministry of Education and Research (BMBF) and the Saxon state government out of the State budget approved by the Saxon State Parliament. 
This work has received funding from the European Union's Just Transition Fund (JTF) within the project \emph{R\"ontgenlaser-Optimierung der Laserfusion} (ROLF), contract number 5086999001, co-financed by the Saxon state government out of the State budget approved by the Saxon State Parliament.
This work has received funding from the European Research Council (ERC) under the European Union’s Horizon 2022 research and innovation programme
(Grant agreement No. 101076233, "PREXTREME"). 
Views and opinions expressed are however those of the authors only and do not necessarily reflect those of the European Union or the European Research Council Executive Agency. Neither the European Union nor the granting authority can be held responsible for them. Computations were performed on a Bull Cluster at the Center for Information Services and High-Performance Computing (ZIH) at Technische Universit\"at Dresden and at the Norddeutscher Verbund f\"ur Hoch- und H\"ochstleistungsrechnen (HLRN) under grant mvp00024.

\bibliography{ref.bib}

\appendix

\section{Effect of Choice of Pseudopotential}

\begin{figure}
    \centering
    \includegraphics[width=\columnwidth,keepaspectratio]{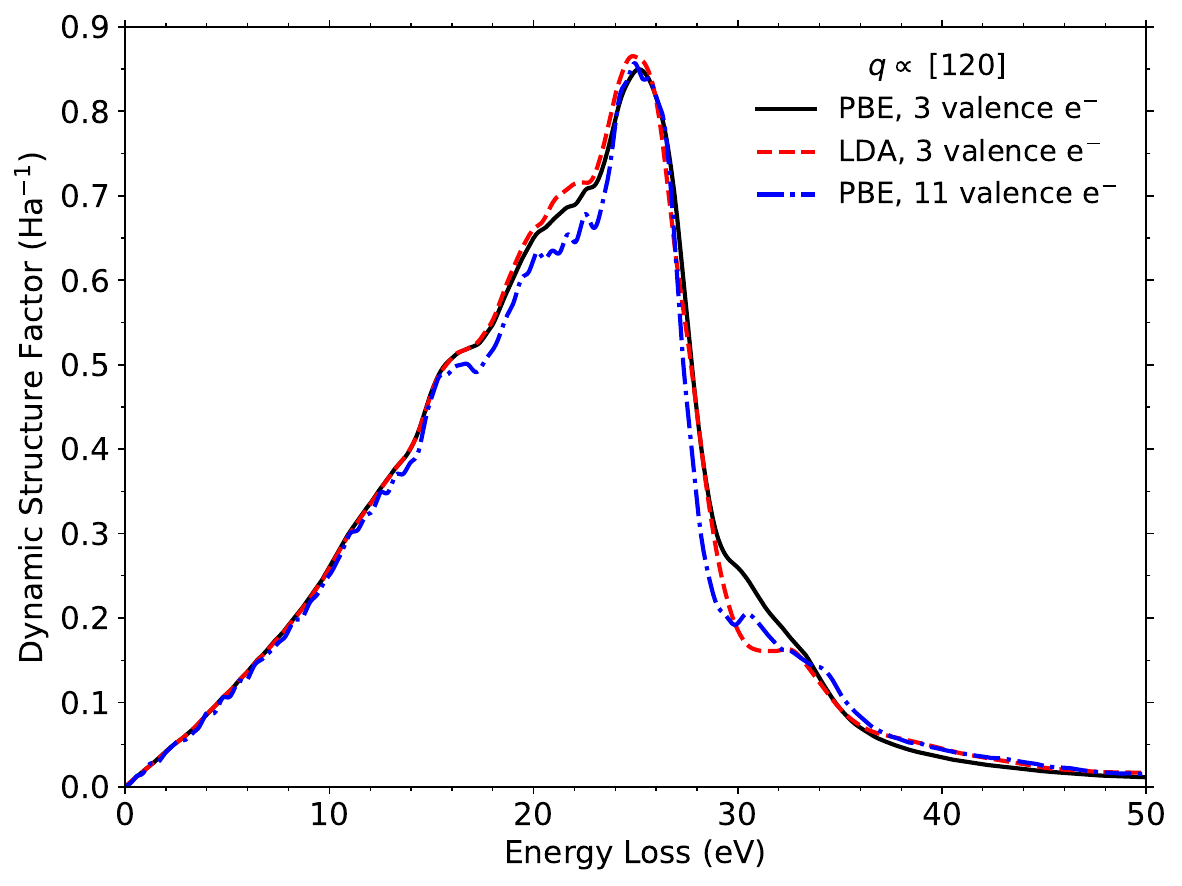}
    \caption{Comparison of different theoretical DSFs along the [120] direction using different pseudopotentials: PBE with 3 valence electrons (black solid; same as main text), LDA with 3 valence electrons (red dashed), and PBE with 11 valence electrons (blue dash-dotted).
    }
    \label{fig:Psuedo}
\end{figure}

The pseudopotential and associated exchange-correlation functional are input parameters into a DFT calculation: the latter due to the exact exchange-correlation function being unknown; and the former to make DFT calculations tractable by reducing the computational cost of modelling the shapes of wavefunctions in the vicinity of the ionic core, as well as freezing some electronic wavefunctions as core states. In constructing a pseudopotential, it needs be optimised to give good predictions of some property of a material, typically atomisation energy or the equation of state~\cite{Lejaeghere_2014_Error}, but not the DSF. 

We therefore compared three different pseudopotentials from the Quantum ESPRESSO pseudopotential database~\cite{PP_ESPRESSO_2022} -- treating either 3 or 11 electrons as valence, and use either PBE or LDA for the exchange-correlation functional -- to see what impact this would have on the DSF. This included: {Al.pbe-rrkj.UPF} (3 valence electrons, PBE, used in main text), {Al.pbe-sp-van.UPF} (11 valence electrons, PBE), and {Al.pz-hgh.UPF} (3 valence electrons, LDA).

A comparisons of the DSFs using these pseudopotentials is shown in Fig.~\ref{fig:Psuedo}. While they have the same general shape, there are some differences depending on the pseudopotential used. The peak positions and heights are similar for all three pseudopotentials. The two 3 valence electron pseudopotentials give similar shapes on the low energy loss wing of the DSF, while the 11 valence electron case is slightly weaker with additional excitations. On the high energy wing between energy losses of 30--32~eV there are some notable differences between the different pseudopotentials. However, once $q$-vector blurring and orientational average is accounted for, we expect these differences to be less prominent between pseudopotentials.
Furthermore, given the most notable discrepancy between the theoretical and experimental curves in Fig.~\ref{fig:Average} is around the maximum, we do not expect selecting a different pseudopotential to improve agreement with experiment much further.

\section{Effect of Lattice Parameter}

\begin{figure}
    \centering
    \includegraphics[width=\columnwidth,keepaspectratio]{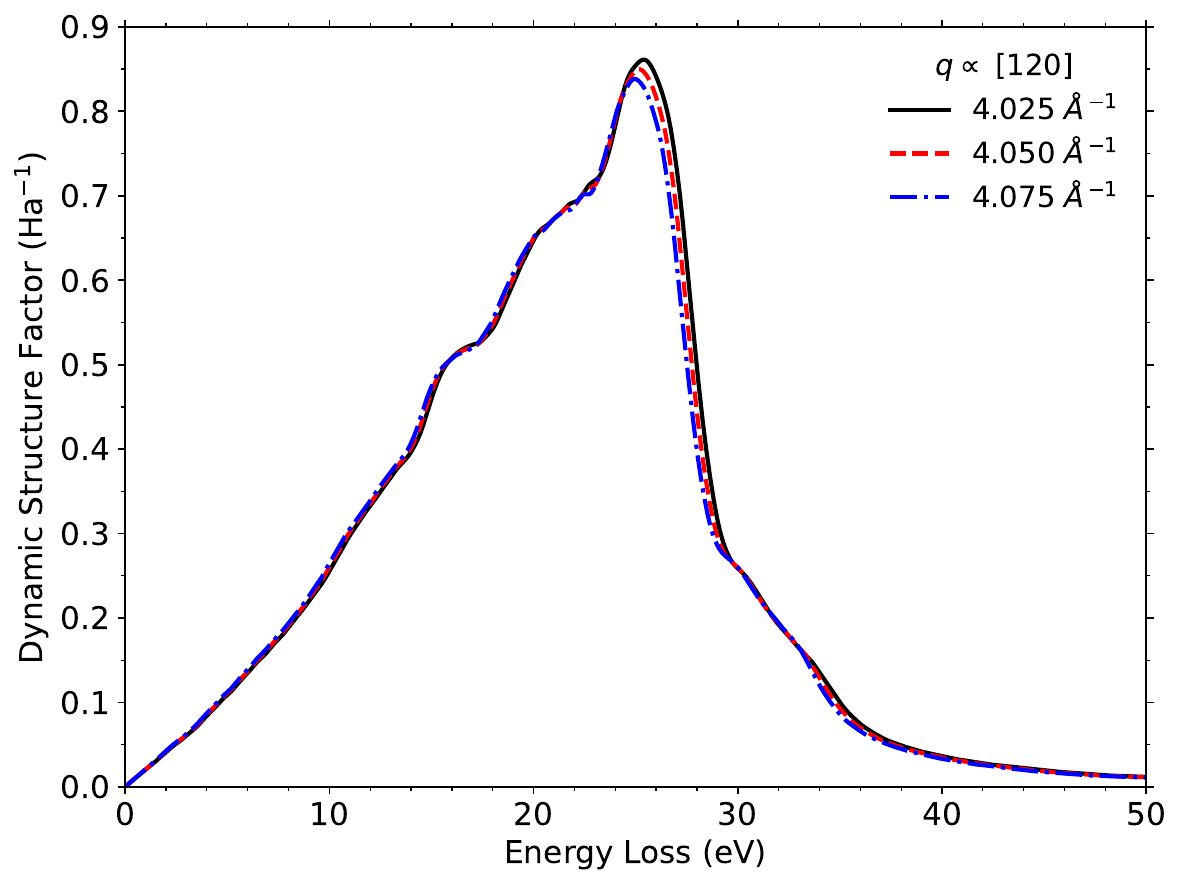}
    \caption{Comparison of different theoretical DSFs along the [120] direction using different lattice parameter sizes: 4.025~$\A^{-1}$ (black solid), 4.050~$\A^{-1}$ (red dashed), and 4.075~$\A^{-1}$ (blue dash-dotted).
    }
    \label{fig:LatticeParm}
\end{figure}

The lattice parameter can also affect the shape of the DSF by changing the electronic density. For different pseudopotentials and XC functionals, the lattice parameter that equilibrates a system can be different~\cite{Haas_2009_Calculation}. Additionally, while an experimental measurement of the lattice parameter of Al is used in the main text, there is also an uncertainty on the exact lattice parameter~\cite{Kittel2004-ta}. To check that the choice of lattice parameter does not strongly affect the results, we compare DSFs for three different lattice parameters in Fig.~\ref{fig:LatticeParm}. Within this range of lattice parameters, the sign of the pressure changes, indicating the equilibrium value for the pseudopotential used in this work lies in this range. Evidently, the shape of the DSF does not strongly depend on the specific lattice parameter. The shift in the maximum of the DSF from the lattice parameter using in the paper ($a=4.05$~$\A^{-1}$) is $+0.22$~eV for $a=4.025$~$\A^{-1}$ and $-0.21$~eV for $a=4.075$~$\A^{-1}$, with the latter shift being much smaller than the +1.45~eV offset of the theoretical maximum position from the experimental maximum.

\end{document}